\magnification=\magstep1
\tolerance=10000

\centerline{{\bf EXACT SOLUTION OF THE ONE-IMPURITY}}
\centerline{{\bf QUANTUM HALL PROBLEM}}
\bigskip
\centerline{{\bf Paola Giacconi}
\footnote{$^{a)}$}{{\tt e-mail: giacconi@bo.infn.it}} and 
{\bf Roberto Soldati}
\footnote{$^{b)}$}{{\tt e-mail: soldati@bo.infn.it}}}
\centerline{{\sl Dipartimento di Fisica "A. Righi", Universit\`a di Bologna,}}
\centerline{{\sl Istituto Nazionale di Fisica Nucleare, Sezione di Bologna,}}
\centerline{{\sl 40126 Bologna - Italy}}
\vskip 2.0 truecm
\centerline{Abstract}
\bigskip
\noindent
{\it The problem of a non-relativistic electron
in the presence of a uniform electromagnetic field
and of one impurity, described by means of an Aharonov-Bohm point-like vortex,
is  studied. The exact solution is found
and the quantum
Hall's conductance turns out to be the same as in the impurity-free case.
This exactly solvable model seems to give indications, concerning the possible 
microscopic mechanisms underlying the integer quantum Hall effect, which
sensibly deviate from some proposals available in the literature.}

\vskip 3.0 truecm
\noindent
To appear in J. Phys. A
\hfill 
June 2000
\bigskip

\vfill\eject
\baselineskip= 22 pt
\noindent 
{\bf 1.\ Introduction.}
\medskip
The discovery of the integer quantum Hall effect [1] represents one of the
most remarkable experimental finding in the last years. The effort towards
an explanation of the experimental plots for the Hall conductance vs electron
density or applied magnetic field has stimulated a huge theoretical activity
[2]. The common key ingredients, among theoretical models that have been
put forward, are the irrelevance of electrons interactions and the central
role played by the presence of impurities within the Hall sample, {\it i.e.},
the effect of disorder. It appears therefrom, that the study of the 
{\it 2+1} dimensional quantum
dynamics of a non-relativistic electron, in the presence of background
electromagnetic fields and of suitable potentials describing disorder, is of
essential importance. Such a simple model should represent the natural starting
point in order to achieve a microscopic description for the integer quantum
Hall phenomenology.

The simplest way to draw one localized impurity could appear to be a
point-like interaction as {\it naively} described by a $\delta$-like potential 
\footnote{$^1$}{In [3] it is
claimed that the model with a single $\delta$-function impurity is
"essentially exactly solvable". However, several approximations and formal
manipulations, which are not mathematically clean, are involved. A rigorous
treatment of the contact interaction in quantum mechanics [4] drives to
different conclusions as we shall see in the sequel.}. 
However, it turns out that a quantum mechanical $\delta$-like potential in two
spatial dimensions is mathematically ill-defined [4] and
what it has to be done properly is to study the most general boundary conditions for the wave functions at the impurity position.
In such a way, conditions of regularity of the wave functions mean absence of point-like or contact interaction,
whereas non-trivial singular boundary conditions of the wave functions at some point entail the presence of a point-like impurity at that point.
To do this, eventually, the
analysis of deficiency indices and subspaces should be worked out in
order to find all the self-adjoint extensions of the corresponding quantum
Hamiltonian, what has not been done insofar in the presence of uniform electric and magnetic fields. 
Actually, a more general and
mathematically consistent way to model one point-like impurity is by means of
a point-like vortex potential [5] of the Aharonov-Bohm type [6] together with 
all its possible self-adjoint extensions, i.e., all the possible boundary conditions at the vortex position. 
In so doing, taking the limit of a vanishing vortex flux, one comes back smoothly to the above case of the pure contact interaction.
On the contrary, the limit of vanishing electric Hall field - which is the one breaking the $O(2)$
symmetry - is very delicate, as it turns out to be entangled with the nature of the domain 
of the quantum Hamiltonian.

Anyway, apart from the
detailed shape of the disorder potentials, the theoretical investigations
concerning the calculation of the density of the quantum states, as well as of
the electric conductivity, actually rely upon approximate methods and, notably,
perturbative approaches [2],[5],[7],[8]. Nonetheless, it turns out - up to
our knowledge - that no exact non-perturbative solutions have been obtained,
even in the simplest realizations of the above mentioned basic
model of the two-dimensional  electron ideal gas in the presence of disorder
potentials.

It is the aim of the present paper to make a first step towards the filling of
that lack, as we shall exhibit and discuss the exact solution for the quantum
mechanical problem of a non-relativistic electron in the presence of a uniform
electromagnetic field and of the Aharonov-Bohm vortex plus contact interaction potential to describe one
impurity.
In spite of its first glance simplicity, the solution of the latter model is
not trivial. As a matter of fact, while in the absence of the electric field
the $O(2)$ rotational symmetry of the model naturally suggests the use of polar
coordinates and of the symmetric gauge, the switching on of the symmetry
breaking uniform electric field does indeed spoil that facility. Consequently,
it appears to be extremely fruitful, in order to find the exact solution, to
follow an algebraic method as well as to employ holomorphic coordinates
[9].
Moreover, to reach our final goal it is necessary to perform some little
mathematical {\it tour de force}, in order to become familiar with the realm
of the self-adjoint extensions of the symmetric radial Hamiltonian operators,
{\it i.e.}, to specify the nature of their domains\footnote{$^2$}{Concerning
definitions and basic theorems on symmetric, self-adjoint and essentially
self-adjoint operators see , {\it e.g.},
N. Akhiezer and I. Glazman, "{\it Theory of Linear Operators in Hilbert
Space}", Frederich Ungar, New York (1963); M. Reed and B. Simon, "{\it
Methods of Modern Mathematical Physics}", vol.2, 
Academic Press (1978).}. 

The present analysis shows that, in the absence of the electric
field, a large - actually infinite - arbitrarity is allowed in the
specification of the quantum radial Hamiltonian operators. On the contrary, 
after the addition of a non-vanishing uniform electric field, the situation
drastically changes: "localized eigenstates"
are no longer allowed in the one-impurity model, all the eigenstates being
improper and non-degenerate - just like in the impurity-free case,
cause the Hamiltonian turns out to be essentially self-adjoint. This is
the ultimate reason why the exact solution is unique. On the other hand, it is
also found that the wave functions of the  improper and non-degenerate
conducting eigenstates are necessarily singular at the impurity position, but in the limiting case of vanishing vortex flux, 
a clear signal that the configuration manifold underlying the model is that of
the one-punctured plane, {\it i.e.} topologically non-trivial.

All those above mentioned features of the model do represent the tools, thanks
to which the total Hall conductance is computed to be the same as in the
"classical" impurity-free problem, according again to the general consensus. However, it should not be missed by the attentive reader that the
detailed quantum mechanical microscopic mechanism, which eventually drives to
the very same current and conductance in the zero- and
one-impurity models, looks to be rather different from the ones usually
acknowledged [2],[8]. In this sense we hope that the exact solution of
the one-impurity model could shed some light on the intimate
microscopic nature leading on the onset of the Hall plateaus.

\bigskip
\noindent 
{\bf 2.\ Exactly solvable models.}
\medskip

In this section we shortly review the exact solutions for the non-relativistic
quantum mechanical motion of a charged point-like particle of charge
$-|e|$ and mass $m$ (one electron) in {\it
2+1} dimensions, first in the presence of uniform - {\it i.e.} constant and
homogeneous - electric and magnetic fields (what is known as the "classical"
Landau's problem) and, second, in the presence of a uniform magnetic field
and of one impurity described by a point-like Aharonov-Bohm vortex . Although
these solutions are very well known, we find it useful to reproduce the results
within the so-called  symmetric gauge and paying special attention to some
algebraic methods that will be quite suitable later on, in order to treat the
one-impurity Hall's problem. In so doing, we also establish our notations and
conventions.

After choosing a uniform magnetic field of strength $B>0$ and orthogonal to
the $Ox_1x_2$ plane in the symmetric gauge
$$
A_j (x_1,x_2)=-\epsilon_{jl}x_l{B\over 2}\ ,\quad
 j,l=1,2\ ,\quad \epsilon_{12}=1\ ,
\eqno(2.1)
$$
we set
$$
\eqalignno{
& z= {x_1+ix_2\over \lambda_B} = x+iy ; & (2.2a)\cr
& x_1=\lambda_B {z+\bar z\over 2}\ ,\ x_2=\lambda_B 
{z-\bar z\over 2i}\ ; & (2.2b)\cr
& \partial_z = (\lambda_B/2)\left(\partial_1-i\partial_2\right)\ , & (2.2c)\cr}
$$
$\lambda_B = \sqrt{\hbar c/|e|B}$ being the magnetic length.
It is now immediate to realize that the
Schr\"odinger-Pauli Hamiltonian for
the spin-up component
\footnote{$^3$}{Throughout this paper we shall always refer to
the spin-up components of the wave functions, 
the generalization to the
spin-down components being straightforward.}
in the presence of an additional uniform electric Hall field of a
strength $E_H>0$, along the positive $Ox_1$ direction, reads
$$
H(E_H)= {\hbar^2\over 2m\lambda_B^2}
\left(2\bar\delta\delta - \varrho{z+\bar z\over 2}\right)\ ,
\eqno(2.3)
$$
where the energy creation distruction operators
$\delta\equiv i\sqrt 2\left\{\partial_{\bar z} + (z/4)\right\} = \bar\delta^\dagger$ and
the dimensionless parameter $\varrho\equiv
2(E_H/B)\sqrt{mc^2/\hbar\omega}$ have been introduced, $\omega\equiv (|e|B/mc)$
being the classical cyclotron's angular frequency. The above Hamiltonian
operator, whose domain is that of the regular wave functions on the plane,
turns out to be self-adjoint since, as we shall see, the eigenvalues are real
and the eigenfunctions span a complete orthonormal set.

Now, there is a nice algebraic way to put the above Hamiltonian into
diagonal form. To this aim, let us first introduce the following set of
translated energy and degeneracy creation-destruction operators respectively:
namely, 
$$
\eqalignno{
& \delta_\varrho \equiv i\sqrt 2\left\{\partial_{\bar z} + 
                {z-\varrho\over 4}\right\} = \bar\delta_\varrho^\dagger\ , 
& (2.4a)\cr
& \bar\delta_\varrho \equiv i\sqrt 2\left\{\partial_{z} - 
                {\bar z -\varrho\over 4}\right\} = \delta_\varrho^\dagger\ ,
& (2.4b)\cr
& \theta_\varrho \equiv -i\sqrt 2\left\{\partial_{z} + 
                {\bar z -\varrho\over 4}\right\} =\bar\theta_\varrho^\dagger\ ,
& (2.4c)\cr
& \bar\theta_\varrho \equiv -i\sqrt 2\left\{\partial_{\bar z} - 
                {z -\varrho\over 4}\right\} = \theta_\varrho^\dagger\ ,
& (2.4d)\cr}
$$
which fulfil the operator algebra
$$
\left[\delta_\varrho ,\bar\delta_\varrho\right]=
\left[\theta_\varrho ,\bar\theta_\varrho\right]=1\ ,\quad
\left[\delta_\varrho ,\theta_\varrho\right]=\left[\delta_\varrho ,
\bar\theta_\varrho\right]=0\ ;
\eqno(2.5)
$$
then,
it is a simple exercise to show that
the Hamiltonian operator (2.3) -  up to the energy scale factor 
$(\hbar^2/2m\lambda_B^2)$ - can be cast into the form
$$
{2m\lambda_B^2\over \hbar^2}H(E_H)\equiv
{\tt h}(\varrho)= 2\bar\delta_\varrho\delta_\varrho
+i{\varrho\over \sqrt 2}(\bar\theta_\varrho -\theta_\varrho) 
                -{3\over 4}\varrho^2\ .
\eqno(2.6)
$$

The above expression for the Hamiltonian, 
together with the operator algebra (2.5), actually suggest that we can search 
for the eigenvectors of ${\tt h}(\varrho)$ as simultaneous eigenstates of the
"Landau-like" Hamiltonian $2\bar\delta_\varrho\delta_\varrho$ 
and of the operator
$$
{\tt T}(\varrho)\equiv i{\varrho\over \sqrt 2}(\bar\theta_\varrho
-\theta_\varrho)-{3\over 4}\varrho^2= -\varrho\left({\lambda_B p_2\over \hbar}
+{x_1-\varrho\lambda_B\over  2\lambda_B}+{3\over 4}\varrho\right)\ ,
\eqno(2.7)
$$
which admits a continuous spectrum and represents the combined effect of a
translation along the $Ox_2$-axis and a gauge transformation. As a matter of
fact, if we introduce the real number $p_\perp\equiv (\lambda_B p_2/\hbar)$
where $p_2$ is the transverse momentum - orthogonal to the electric field - 
we obtain
$$
{\tt T}(\varrho)
\exp\left\{iy\left(p_\perp -{x\over 2}\right)\right\}=
\left(-\varrho p_\perp -{1\over 4}\varrho^2\right)
\exp\left\{iy\left(p_\perp -{x\over 2}\right)\right\}\ .
\eqno(2.8)
$$

It follows therefrom, that the Hamiltonian (2.6) has a continuous
non-degenerate spectrum whose eigenvalues are given by
$$
\varepsilon_{n,p_\perp}= 2n -\varrho p_\perp -{1\over 4}\varrho^2\ ,\qquad
n+1\in{\bf N}\ ,\quad p_\perp\in {\bf R}\ ,
\eqno(2.9)
$$
which reproduce the well known electric field splitting of the Landau's
bands.

The eigenfunctions can be written in the following forms: namely,
$$
\eqalign{
\left<xy|\psi_{n,p_\perp}(\varrho)\right>&\equiv\psi_{n,p_\perp}(x,y;\varrho)
 = \sum_{k=0}^\infty
c_k^{(n)}(p_\perp)\varphi_{n,k}(z,\bar z;\varrho)\cr
& =
u_n\left(x-{1\over 2}\varrho -p_\perp\right)
\exp\left\{iyp_\perp -i{xy\over 2}\right\}\sqrt{{1\over 2\pi}}\ ,\cr}
\eqno(2.10)
$$
in which
$$
c_k^{(n)}(\tilde p)= (-i)^k u_k (\tilde p)\ ,\qquad k+1,n+1\in{\bf N}\ ,
\eqno(2.11)
$$
where $\left\{u_k,\ k+1\in{\bf N}\right\}$ is the complete
orthonormal set of the Hermite's functions and
$$
\eqalign{
& \left<z\bar z|n,k;\varrho\right>\equiv \varphi_{n,k}(z,\bar z;\varrho)=
\sqrt{{1\over 2\pi}}\exp\left\{-{1\over 4}(\bar z -\varrho)(z- 
\varrho)\right\}\times\cr
& \left(-i{\bar z -\varrho\over \sqrt 2}\right)^n
\left(i{z -\varrho\over \sqrt 2}\right)^k
\sum_{h=0}^\infty \left(-{2\over |z-\varrho|^2}\right)^h
{\sqrt{n!k!}\over h!\Gamma (n+1-h)\Gamma (k+1-h)}\ .\cr}
\eqno(2.12)
$$
is the complete orthonormal Bargmann-Segal set.
It readily follows therefrom that the 
improper eigenfunctions (2.10) are complete and orthonormalized in the 
continuum as
$$
\left<\psi_{n,p_\perp}(\varrho)|\psi_{m,q_\perp}(\varrho)\right>=\delta_{n,m}\delta
(p_\perp - q_\perp)\ ,\qquad p_\perp,q_\perp\in {\bf R}\ , n+1,m+1\in{\bf N}\ .
\eqno(2.13)
$$

Finally, we briefly recall that
the "classical" Hall's conductance of each eigenstate is provided
by
$$
\sigma_{xy} = -{e^2\over h}\Gamma_L^{-1}\ ,
\eqno(2.14)
$$
where the usual Landau's levels degeneracy factor is 
$$
\Gamma_L\equiv {1\over 2\pi\lambda_B^2}={|e|B\over hc}\ .
\eqno(2.15)
$$
According to the above described simple property, it turns out that the total
Hall conductance of an ideal electron gas in a pure sample
is proportional to the filling factor $\nu=({\tt n}/\Gamma_L)$,  where ${\tt
n}$ denotes the number of electrons {\it per} unit area.
\medskip
The problem of a point-like charged particle on the plane 
in the presence of a uniform magnetic field and one Aharonov-Bohm 
point-like singularity - the AB-vortex - has been already solved
in the literature [10]. By the way, it turns out that the AB-vortex
faithfully describes [5] the presence of some localized impurity within 
the Hall's sample.
It is quite instructive to resolve this problem
by means of a suitable algebraic method. In so doing, in fact, it is possible
to unravel some interesting features of the exact solutions, which have been
not yet discussed insofar, up to our knowledge, but will be crucial in order
to provide the exact solution in the presence of an additional uniform electric
field.  

The gauge potential, in the symmetric gauge, is now given by
$$
A_j(x_1,x_2)=-\epsilon_{jl}x_l\left({B\over 2}-
{(\phi/2\pi)\over x_1^2+x_2^2}\right)\ ,
\eqno(2.16)
$$
in which the flux-parameter $\phi>0 (<0)$ means that the vortex magnetic 
field, located at the origin,
is anti-parallel (parallel) to the uniform magnetic field $B>0$. After 
introduction of the quantum flux unity $\phi_0\equiv (hc/|e|)$ and of the
dimensionless parameter $\alpha\equiv (\phi/\phi_0)$, it can be easily shown
that the rescaled Schr\"odinger-Pauli Hamiltonian for the upper spinor
component takes the form
$$
{\tt h}(\alpha)= 2\bar\delta (\alpha)\delta (\alpha)\ ,
\eqno(2.17)
$$
where the singular creation-destruction energy operators appear to be
$$
\eqalignno{
& \delta (\alpha)\equiv i\sqrt 2\left\{\partial_{\bar z} + {z\over 4}
\left(1-{\alpha\over [\gamma]}\right)\right\}=
\bar\delta^\dagger(\alpha)\ , & (2.18a)\cr
& \bar\delta (\alpha)\equiv i\sqrt 2\left\{\partial_z - {\bar z\over 4}
\left(1-{\alpha\over [\gamma]}\right)\right\}=
\delta^\dagger (\alpha)\ ,    & (2.18b)\cr}
$$
with $\gamma\equiv (\bar z z/2)$. The singularity at $\gamma =0$ in the
foregoing expressions is understood in the sense of the tempered
distributions [11]: namely,
$$
\eqalign{
{1\over [\gamma]} & \equiv \left({\partial^2\over \partial x^2}+
{\partial^2\over \partial y^2}\right)\left(\ln\sqrt{x^2+y^2}\right)^2
+C\delta (x)\delta (y)\cr
& = {1\over 4}\triangle\ln^2 (\bar z z) + C\delta^{(2)}(\bar z,z)\ ,\cr}
\eqno(2.19)
$$
where $C$ is an arbitrary constant, whose presence ensures {\it naive}
scaling behavior of the tempered distribution itself, {\it i.e.}, $1/[c\gamma]=
(1/c)(1/[\gamma]),\ c>0$. To
start with, let us consider the domain of the operators (2.18) to be the
Besov's space  ${\cal T}({\bf R}^2)=\{f\in {\cal S}({\bf R}^2)|\ f(0)=0\}$,
which is dense in $L^2({\bf R}^2)$.

To be definite and without loss of generality, we shall choose in the sequel  
$-1<\alpha< 0$, corresponding to parallel uniform and vortex magnetic
fields. As a matter of fact, it is well known that only the non-integer part
of $\alpha$ is relevant, its integer part being always reabsorbed by
means of a single-valued gauge transformation.

In order to fully solve the eigenvalue problem, it is convenient to introduce
also the associated singular creation-destruction degeneracy operators
$$
\eqalignno{
& \theta (\alpha)\equiv -i\sqrt 2\left\{\partial_{z} + {\bar z\over 4}
\left(1+{\alpha\over [\gamma]}\right)\right\}=
\bar\theta^\dagger (\alpha)\ , & (2.20a)\cr
& \bar\theta (\alpha)\equiv -i\sqrt 2\left\{\partial_{\bar z} - {z\over 4}
\left(1+{\alpha\over [\gamma]}\right)\right\}=
\theta^\dagger (\alpha)\ , & (2.20b)\cr}
$$
always acting on the same domain ${\cal T}({\bf R}^2)$, in such a way
that the following simple operator algebra still holds true for any
$-1<\alpha\le 0$: namely,
$$
[\delta (\alpha),\bar\delta (\alpha)]=[\theta (\alpha),\bar\theta (\alpha)]=
1\ ,\quad [\delta (\alpha),\theta (\alpha)]=
[\delta (\alpha),\bar\theta (\alpha)]=0\ .
\eqno(2.21)
$$
We notice that, in order to reproduce the foregoing algebra, it is essential 
to employ the definition (2.19). As a matter of fact, eq.~(2.19) guarantees
the  {\it naive} action of the dilatation operator 
$$
{\tt D}{1\over [\gamma]}= -2{1\over [\gamma]}\ ,\quad
{\tt D}\equiv z\partial_z +\bar z\partial_{\bar z}\ ,
\eqno(2.22)
$$
whence it is an easy exercise to check the algebra (2.21).

Now, owing to $O(2)$-symmetry of the problem, we can search for common
eigenfunctions of the rescaled Hamiltonian (2.17) and of the angular
momentum operator
$$
{\tt L}\equiv \hbar(z\partial_z -\bar z\partial_{\bar z})=
\hbar\{|\alpha|{\bf 1}+\bar\theta (\alpha)\theta (\alpha)-\bar\delta (\alpha)
\delta (\alpha)\}\ ,
\eqno(2.23)
$$
which manifestly commutes with ${\tt h}(\alpha)$.

Let us consider in the present section the case in which the domain of the
rescaled Hamiltonian (2.17) is ${\cal S}({\bf R}^2)$ - wave functions regular
at the origin - which is dense in $L^2({\bf R}^2)$ and provides the standard
solution given in the literature [10].
The eigenstates are naturally separated into two classes named
integer-valued energy eigenstates
(I.V.E.), which form an infinite degenerate set,
and real-valued energy eigenstates 
(R.V.E.) whose degeneracy is always finite. 
Now, it turns out that the $n$-th Landau's band 
of rescaled energy $\check\varepsilon_n=2n$ is spanned by the I.V.E.
eigenstates 
$$
\eqalign{
\left<z\bar z|n<k;\check\alpha\right> & =
(-1)^n\sqrt{{n!\over 2\pi\Gamma (k+\alpha +1)}}
\left(i{z\over \sqrt 2}\right)^{k-n}\gamma^{-|\alpha|/2}\exp\{-\gamma /2\}
L_n^{(k-n+\alpha)}(\gamma)\cr
& \equiv \Psi_{n<k}(z,\bar z)\ ,\qquad
 k\ge n+1\in{\bf N}\ ,\cr}
\eqno(2.24)
$$
$L_n^{(\beta)}$ being the generalized Laguerre's
polynomials. It is worthwhile to remark that the set of the above eigenstates
(2.24) actually realizes the
infinite degeneracy of the Landau's bands, 
the degeneracy being labelled by
the quantum number $k\ge n+1$. Notice also that the integer-valued energy bands
contain an infinite number of states, 
although $n+1$ states
less than the corresponding ordinary Landau's band in the absence of
the AB-vortex impurity. Finally, all the I.V.E. actually belong to
${\cal T}({\bf R}^2)$, since it is easy to check that 
their holomorphic representations
do vanish at the origin cause, as it appears to be manifest from the above
expression, the items $k=0,1,2,\ldots ,n$ are forbidden as they drive
outside either the domain of the Hamiltonian ($k=n$) or even 
outside
$L^2({\bf R}^2)$ ($k=0,1,\ldots ,n-1$).

The (R.V.E.) eigenstates correspond to non-integer eigenvalues
$\hat\varepsilon_n=2(n+|\alpha|)$ of the rescaled Hamiltonian,
the corresponding eigenfunctions being
$$
\eqalign{
\left<z\bar z|n\ge k;\hat\alpha\right> & =
(-1)^k\sqrt{{k!\over 2\pi\Gamma (n-\alpha +1)}}
\left(-i{\bar z\over \sqrt 2}\right)^{n-k}\gamma^{|\alpha|/2}\exp\{-\gamma /2\}
L_k^{(n-k-\alpha )}(\gamma)\cr
& \equiv \Psi_{n\ge k}(z,\bar z)\ ,\quad n+1\in {\bf N}\ ,\quad
 0\le k\le n\ ,\cr}
\eqno(2.25)
$$
which, again, belong to the above specified domain of ${\tt h}(\alpha )$
iff  the degeneracy quantum number $k$ does not exceed the energy quantum
number $n$.

\bigskip
\noindent
{\bf 3.\ Self-adjoint extensions of the Hamiltonian.}
\medskip

We have considered insofar the rescaled Hamiltonian ${\tt h}(\alpha)$
to be defined on the domain of the regular square-integrable wave
functions. However, as it is well known, this is not the most general case.
Let us now consider, therefore, different quantum Hamiltonians, corresponding
to different self-adjoint Hamiltonian operators, whose differential operator
is always given by eq.~(2.17), but whose domain is now allowed to contain
wave functions with square integrable singularities at the origin -
{\it i.e.} at the AB-vortex position.
This procedure is the mathematically correct way to introduce in the context
contact-interaction  or point-like interaction. In physical
terms, it is equivalent to  {\it naively} add some kind of $\delta$-like
potential to the classical Hamiltonian. We have to stress in fact that,
strictly speaking, $\delta$-like potential are ill-defined in two and three
spatial  dimensions [4],[13]  and the proper way to encompass the possibility
of contact-interaction is by means of the analysis of the self-adjoint
extensions of the quantum Hamiltonian. In particular, the solution we have
discussed in the previous section, {\it i.e.} the case of the Hamiltonian
whose domain is that of the regular wave functions, can be thought of as the
pure AB interaction in the absence of contact-interaction. The presence of a
particular square integrable singularity of the wave funtion at the vortex
position  will select some new quantum Hamiltonian, which will describe the
presence of a specific contact-interaction.
What we shall see in the sequel is that there is an
infinite number of such Hamiltonians, which are perfectly legitimate and turn
out to describe different physics, as they are characterized by different
spectra and  degeneracies. As a matter of fact, it is not difficult to prove
the following
\medskip
{\sl Lemma}\ $(-)$:
\par\noindent
in any subspace of fixed negative or vanishing angular
momentum $\ell = -\hbar l\ ,\ l+1\in {\bf N}$ there are
two options in order to specify the quantum radial Hamiltonian:
if the domain is that of regular wave functions we have
$$
{\tt h}_l (\alpha) =\sum_{n=l}^\infty 2(n+|\alpha|)\hat P_{n>n-l}
(\alpha)\ ,
\eqno(3.1)
$$
where the projectors onto the regular R.V.E. eigenstates of eq.~(2.25) are
introduced, {\it i.e.}, $\hat P_{n\ge k}(\alpha)\equiv\left|n\ge
k;\hat\alpha\right> \left<n\ge k;\hat\alpha\right|$.
Alternatively,
if the domain is that of the wave functions which are square integrable on 
the plane, although singular at the impurity's position, we have
$$
{\tt H}_l (\alpha) =\sum_{n=l}^\infty 2n\check P_{n\ge n-l}
(\alpha)\ ,
\eqno(3.2)
$$
where 
$$
\check P_{n\ge n-l}(\alpha)\equiv \left|n\ge n-l;\check\alpha\right>
\left<n\ge n-l;\check\alpha\right|\ ,
\eqno(3.3)
$$
the state $\left|n\ge n-l;\check\alpha\right>$ being given by
$$ 
\left<z\bar z|n\ge n-l;\check\alpha\right>\equiv\Phi_{n>n-l} (z,\bar z)=
\sum_{j=0}^\infty \check c_{n,l}^{\ j} \hat\psi_{j,j-l}(z,\bar z ;\alpha)\ ,
\qquad l=0,1,\ldots ,n\ ,
\eqno(3.4)
$$
where
$$
\eqalign{
\check c_{n,l}^{\ j} &\equiv
\left<j>j-l;\hat\alpha|n>n-l;\check\alpha\right>\cr &=
{j-\alpha\choose n}{n+\alpha-l\choose j-l}
\sqrt{{(j-l)!n!\over \Gamma(j-\alpha +1)\Gamma (n-l+\alpha +1)}} \ .\cr}
\eqno(3.5)
$$
The above eq.~(3.5) uniquely defines a state vector $\forall n+1\in {\bf N}$,
owing to the Riesz-Fisher theorem, since it can be actually verified that
$$
\eqalign{
\sum_{k=0}^\infty \left|\check c_{n,l}^{\ k}\right|^2 &=
{\left\{(n+\alpha)(\alpha)_n\right\}^2\over
n!\Gamma(1-\alpha)\Gamma(n-l+\alpha+1)[(-n-\alpha)_l]^2}\cr
&\times
\sum_{k=0}^\infty{(1-\alpha)_k\over (k-l)!(n+\alpha-k)^2}=1\ .\cr}
\eqno(3.6)
$$ 
Notice that the wave functions (3.4) belong to $L^2({\bf R}^2)$ by 
construction, they are singular at the AB-vortex
position and are eigenfunctions of the symmetric operator (2.17) with
integer eigenvalues $\check\varepsilon_n=2n$ and of the angular momentum
operator (2.23) with a negative or vanishing eigenvalue $\ell = -\hbar l$.
The net result of this construction is that, by relaxing the 
condition of the regularity of the wave functions at the AB-vortex position 
- which specifies a particular domain of the quantum Hamiltonian - it
is possible to set up different quantum Hamiltonians, with different
spectra and degeneracies, after shifting an infinite set of states
(actually orthonormal and complete in the subspaces of fixed angular
momenta) from the R.V.E. sector of the eigenstates to the I.V.E. one. 

A quite similar construction can be done for the quantum radial
Hamiltonians corresponding to positive angular momenta $\ell =\hbar
l,\ l\in {\bf N}$. Again, the result can be summarized within the following
\medskip 
{\sl Lemma}\ $(+)$:
\par\noindent
in any subspace of fixed positive angular
momentum $\ell = \hbar l\ ,\ l\in {\bf N}$ there are
two options in order to specify the quantum radial Hamiltonian:
if the domain is that of regular wave functions we have
$$
{\tt h}_l (\alpha) =\sum_{n=l}^\infty 2n\check P_{n<n+l}
(\alpha)\ ,
\eqno(3.7)
$$
in which the projectors onto the regular I.V.E. eigenstates of eq.~(2.24) are
introduced. Alternatively,
if the domain is that of the wave functions which are square integrable on 
the plane, although singular at the impurity's position, we have
$$
{\tt H}_l (\alpha) =\sum_{n=l}^\infty 2(n+|\alpha|)\hat P_{n<n+l}
(\alpha)\ ,
\eqno(3.8)
$$
where 
$
\hat P_{n<n+l}(\alpha)\equiv \left|n<n+l;\hat\alpha\right>
\left<n<n+l;\hat\alpha\right|\ ,
$
the singular state $\left|n<n+l;\hat\alpha\right>$ being given by
$$ 
\left<z\bar z|n<n+l;\hat\alpha\right>\equiv\Phi_{n<n+l} (z,\bar z)=
\sum_{j=0}^\infty \hat c_{n,l}^{\ j} \check\psi_{j,j+l}(z,\bar z ;\alpha)\ ,
\qquad l\in{\bf N}\ ,
\eqno(3.9)
$$
where
$$
\eqalign{
\hat c_{n,l}^{\ j} &\equiv \left<j<j+l;\check\alpha|n<n+l;\hat\alpha\right>\cr
&={j+\alpha+l\choose n+l}{n-\alpha\choose j}
\sqrt{{j!(n+l)!\over \Gamma(j+l+\alpha+1)\Gamma (n-\alpha +1)}} \ .\cr}
\eqno(3.10)
$$
Again, it can be readily verified that
$$
\sum_{j=0}^\infty \left|\hat c_{n,l}^{\ j}\right|^2=
{(n-\alpha)^2\left\{(-\alpha)_n (1+\alpha)_l\right\}^2\over
(n+l)!\Gamma(n-\alpha+1)\Gamma(\alpha+l+1)}
\sum_{k=0}^\infty {(\alpha+l+1)_k\over k!(n-\alpha-k)^2}=1\ ,
\eqno(3.11)
$$
which means that, according to the Riesz-Fisher theorem, the expansion (3.9)
uniquely defines a state vector in the Hilbert space.

It is crucial to gather that the action of the lowering and raising
energy and degeneracy  operators, which were originally defined on ${\cal
T}({\bf R}^2)$, can be extended on the singular states in terms of their
$L^2$-expansions, {\it e.g.}, $$
\eqalign{
&\theta(\alpha)\left|n\ge n-l;\check\alpha\right> 
\equiv \sum_{k=0}^\infty\check c_{n,l}^{\ k}\ \theta(\alpha) 
\left|k\ge k-l;\hat\alpha\right>=\cr
&\sum_{k=1}^\infty\check c_{n,l}^{\ k} 
\sqrt{k-l}\left|k> k-l-1;\hat\alpha\right>=
\sqrt{n-l+\alpha}\left|n> n-l-1;\check\alpha\right>\ , \cr}
\eqno(3.12)
$$
and analogous ones for the remaining raising and lowering operators.

Concerning self-adjoint extensions, we should go a little bit further in
order to reach the most general statement. Actually, it can be proved that,
for any fixed value $\ell=\hbar l,\ l\in{\bf Z}$ of the angular momentum,
there is a continuous family of self-adjoint extensions of the symmetric radial
Hamiltonians which interpolates between the regular one $h_l(\alpha)$ and the
singular one $H_l(\alpha)$. However, since none of those further possible
self-adjoint extensions will be relevant in searching an exact solution of the
one impurity problem in the presence of the uniform electric field, we shall no
longer discuss here that quite interesting matter, but leave it to a 
forthcoming analysis. To sum up, we can say that the solution we have
discussed in the previous section - in which the domain of the quantum
Hamiltonian is that of the regular wave functions - actually corresponds to
one-impurity described by a pure AB interaction. The further possible choices
of the quantum self-adjoint Hamiltonians - such that the domains contain
singular wave functions at the vortex position - do physically represent the
simultaneous presence of the AB and contact-interactions.

\bigskip
\noindent
{\bf 4.\ Exact solution for the one-impurity quantum Hall's problem.}
\medskip
We are now ready to discuss the exact solution in the presence of
the AB-vortex - the one-impurity problem - and of a uniform electromagnetic
field. According to the conventional picture [2] it is conjectured
that the presence of a not too large number of localized impurities within the
Hall's sample is actually what is needed to give account for the onset of
the Hall's plateaus. To this respect, it is plausibly believed that the 
structure in terms of Landau's bands is basically kept, even in the presence
of a small number of localized impurities, although the density of the 
states among and within the Landau's subbands is significantly changed by
the presence of impurities (Landau's subbands are broadened and depopulated).
It is commonly accepted that the above pattern eventually supports, in terms of
various analytical approximate methods of investigations [2], some reasonable
explanation for the Hall's plateaus. The basic idea behind this picture is
that the switching on of a weak electric field is a smooth perturbation whose
net effect is, on the one hand, to lift the degeneracy of the
conducting depopulated Landau's bands whereas, on the other hand, 
to allow the presence of non-conducting localized eigenstates, {\it i.e.},
bound states. 

On the contrary, as we shall see below, the exact solution of the present
model shows that the switching on of the uniform Hall field $E_H$ drastically
modify the distribution and the nature of the energy eigenstates, with respect
to the situation in the absence of $E_H$ and no matter how weak the Hall field
is. In particular, all the energy eigenstates are improper, each of them does
contribute to the Hall's current and, moreover, the improper ({\it
i.e.} extended) wave functions of some of the eigenstates necessarily become
singular at the impurity position.   

In order to prove the above statements, we shall solve our problem 
following a constructive approach, which makes use of all the detailed explicit
information we have learned in the previous sections. As a matter of fact,
what we have seen before is that the presence of the
one-impurity AB-vortex might indeed realize what is widely believed: the
integer valued Landau's levels are kept and, in general, further non-integer
valued energy levels do actually appear, in such a way that the I.V.E.
eigenstates degeneracy of the Landau's band is lowered, albeit still
infinite. This means, in turn, that the density of the states is actually
sensibly modified by the presence of the impurity. Consequently, what is
reasonably expected - following the afore mentioned popular belief - after
switching on some weak uniform electric Hall field, is that the  conductance
of the remaining charged states within the  Landau's bands is slightly
increased - with respect to the  impurity-free case - in such a way that the
net result for the Hall's conductance is again the "classical" one  of
eq.~(2.14). As we will see below, it turns out that the exact solution
actually suggests some quite different picture.

It is not difficult to verify that the rescaled Hamiltonian differential
operator, in the presence of an additional uniform electric field suitably
described by the afore introduced parameter $\varrho$ - see eq.~(2.3) - can be
written in the form
$$
{2m\lambda_B^2\over \hbar^2}H(\alpha ,E_H)\equiv
{\tt h}(\alpha ,\varrho)= 2\bar\delta_\varrho (\alpha )\delta_\varrho
(\alpha )
+i{\varrho\over \sqrt 2}[\bar\theta_\varrho (\alpha )-\theta_\varrho 
(\alpha )] -{3\over 4}\varrho^2\ ,
\eqno(4.1)
$$
in which the translated energy and degeneracy creation-annihilation
operators are referred to be respectively
$$
\eqalignno{
& \delta_\varrho (\alpha)\equiv i\sqrt 2\left\{\partial_{\bar z} + {z\over 4}
\left(1-{\alpha\over [\gamma]}\right) -{\varrho\over 4}\right\}=
\bar\delta_\varrho^\dagger(\alpha)\ , & (4.2a)\cr
& \bar\delta_\varrho (\alpha)\equiv i\sqrt 2\left\{\partial_z - {\bar z\over 4}
\left(1-{\alpha\over [\gamma]}\right)+{\varrho\over 4}\right\}=
\delta^\dagger_\varrho (\alpha)\ ,    & (4.2b)\cr
& \theta_\varrho (\alpha)\equiv -i\sqrt 2\left\{\partial_{z} + {\bar z\over 4}
\left(1+{\alpha\over [\gamma]}\right) -{\varrho\over 4}\right\}=
\bar\theta^\dagger_\varrho (\alpha)\ ,& (4.2c)\cr
& \bar\theta_\varrho (\alpha)\equiv -i\sqrt 2\left\{\partial_{\bar z} - 
{z\over 4}
\left(1+{\alpha\over [\gamma]}\right) + {\varrho\over 4}\right\}=
\theta^\dagger_\varrho (\alpha)\ .    & (4.2d)\cr}
$$
Again, the following canonical commutation
relations hold true: namely,
$$
[\delta_\varrho (\alpha),\bar\delta_\varrho (\alpha)]=[\theta_\varrho
(\alpha),\bar\theta_\varrho (\alpha)]= 1\ ,\quad [\delta_\varrho
(\alpha),\theta_\varrho (\alpha)]= [\delta_\varrho (\alpha),\bar\theta_\varrho
(\alpha)]=0\ . 
\eqno(4.3)
$$

Now, owing to the above algebra, we have that the full Hamiltonian
differential operator ${\tt h}(\alpha,\varrho)$ and the translated
"Landau-like" differential operator 
$2\bar\delta_\varrho(\alpha)\delta_\varrho(\alpha)$ indeed commute, {\it i.e.},
$$
\left[{\tt h}(\alpha,\varrho),
\bar\delta_\varrho(\alpha)\delta_\varrho(\alpha)\right]=0\ . 
\eqno(4.4) 
$$
It is important to gather that, unless we specify the domains of the above
mentioned Hamiltonian differential operators, they are only symmetric.
Since we have to deal with well defined self-adjoint quantum Hamiltonians, we
have  to specify the (common) domain in which the commutation relation (4.4)
still holds for the corresponding quantum self-adjoint Hamiltonians.
But then, the fundamental theorem states that the self-adjoint realizations of
the full rescaled Hamiltonian and of the translated "Landau-like" Hamiltonian
must have a complete orthonormal set of common eigenstates.

First we prove that there is only one choice of the domain of the quantum
Hamiltonians which allows for a solution of the problem. As a matter of fact,
it appears that the spectrum of any self-adjoint extension ${\tt
H}(\alpha,\varrho)$ of the full rescaled Hamiltonian (4.1) is continuous owing
to the presence of the degeneracy lifting operator 
$$
{\tt T}(\alpha,\varrho)\equiv i{\varrho\over
\sqrt 2}[\bar\theta_\varrho (\alpha )-\theta_\varrho  (\alpha )] -{3\over
4}\varrho^2\ ,
\eqno(4.5)
$$
whose spectrum is manifestly continuous - in the "classical"
impurity-free case it drives to the electric splitting of the Landau's
degeneracy, see eq.~(2.9). Consequently the eigenfunctions of ${\tt
H}(\alpha,\varrho)$  will be improper state vectors and must be, owing to
$[{\tt H}(\alpha,\varrho),\Delta_L(\alpha,\varrho)]=0$,
common eigenstates of the corresponding
self-adjoint extension 
$\Delta_L(\alpha,\varrho)$ and of the translated "Landau-like" Hamiltonian
$2 \bar\delta_\varrho(\alpha)\delta_\varrho(\alpha)$, whose spectrum is
instead purely discrete. As a consequence, discrete energy levels of
$\Delta_L(\alpha,\varrho)$ with a finite degeneracy are forbidden, cause
the degenerate states are proper state vectors and a finite combination of
them cannot produce an improper state. This means, in particular, that if
we choose the domain to be, {\it e.g.}, that of the regular wave functions on
the plane, then the full rescaled Hamiltonian (4.1) is not a self-adjoint
operator.

This quite general and rigorous result is such a stringent constraint that we
are left with only two possible options in order to obtain a solution, namely
we have to investigate the two self-adjoint extensions of the translated
"Landau-like" Hamiltonian whose spectra are given by either non-integer
rescaled eigenvalues $\hat\varepsilon_n=2(n+|\alpha|),\ n+1\in{\bf N}$, or,
alternatively, by integer rescaled eigenvalues  $\check\varepsilon_n=2n,\
n+1\in{\bf N}$.   

In the former case, the self-adjoint translated "Landau-like" Hamiltonian is
given by its spectral decomposition: namely,
$$
\hat\Delta(\alpha,\varrho)\equiv\sum_{n=0}^\infty
2(n+|\alpha|)\left\{\sum_{k=0}^n\hat P_{n\ge k}(\alpha,\varrho)
+\sum_{k=n+1}^\infty\hat P_{n<k}(\alpha,\varrho)\right\}\ ,
\eqno(4.6)
$$
where the projectors onto translated regular and singular states are
given by, respectively, 
$$
\eqalignno{
\hat P_{n\ge k}(\alpha,\varrho)&\equiv \left|n\ge k;\hat\alpha,\varrho\right>
\left<n\ge k;\hat\alpha,\varrho\right|\ ,
&(4.7a)\cr
\hat P_{n< k}(\alpha,\varrho)&\equiv \left|n<k;\hat\alpha,\varrho\right>
\left<n<k;\hat\alpha,\varrho\right|\ .
&(4.7b)\cr}
$$

The explicit form of the above eigenstates, normalized to unity, is provided
according to the general construction described in the previous section - see
{\it Lemmas} ($\pm$) - {\it i.e.}, 
$$ 
\left|n,k;\hat\alpha,\varrho\right>=\sqrt{{\Gamma
(1-\alpha)\over (k!)\Gamma (n+1-\alpha)}}[\bar\delta_\varrho(\alpha)]^n
[\bar\theta_\varrho (\alpha )]^{k} \left|0,0;\hat\alpha,\varrho\right>\ ,\quad
n+1, k+1\in{\bf N}\ , \eqno(4.8)
$$
the holomorphic representation of the cyclic ground state being
$$
\left<z\bar z|0,0;\hat\alpha,\varrho\right>=
{\gamma^{|\alpha|/2}\exp\{-(1/4)(z-\varrho)(\bar z-\varrho)\}\over
\sqrt{2\pi\Gamma(1-\alpha)}}{\exp\{-\varrho^2/4\}\over
\sqrt{_1F_1(1-\alpha,1;\varrho^2/2)}}\ .
\eqno(4.9)
$$
Notice that, among the eigenstates (4.8), the regular ones correspond to
negative or vanishing angular momenta ($n\ge k$), whilst the singular ones
to positive angular momenta ($n<k$). Furthermore, it is manifest from the
spectral decomposition (4.6) that the quantum number $k$ labels the infinite
discrete Landau's degeneracy. It can be readily verified, taking the
construction leading to eq.~(3.12) suitably into account, that the following
properties hold true: namely, 
$$
\eqalignno{
&\theta_\varrho(\alpha)\left|n,k;\hat\alpha,\varrho\right>=
\sqrt{k}\left|n,k-1;\hat\alpha,\varrho\right>\ , &(4.10a)\cr
&\bar\theta_\varrho(\alpha)\left|n,k;\hat\alpha,\varrho\right>=
\sqrt{k+1}\left|n,k+1;\hat\alpha,\varrho\right>\ , &(4.10b)\cr}
$$

Now, in order to find the solution of the eigenvalue problem for the quantum
self-adjoint Hamiltonian 
$$
\hat H(\alpha,\varrho)=\hat\Delta(\alpha,\varrho)+{\tt T(\alpha,\varrho)}\ ,
\eqno(4.11)
$$
let us consider the states
$$
\eqalign{
&\left|n,p_\perp;\hat\alpha,\varrho\right>\equiv
\sum_{k=0}^\infty
c_k^{(n)}(\tilde p)\left|n,k;\hat\alpha,\varrho\right>\ ,\cr
&\tilde p\equiv p_\perp-{1\over
2}\varrho\ ,\quad p_\perp\in {\bf R}\ ,\cr} 
\eqno(4.12) 
$$
which are build up in close analogy with the "classical" solution (2.10) -
with $c_k^{(n)}(\tilde p)$ given by eq.~(2.11) -
and belong by definition to the continuous spectrum. Notice that, by
construction, the above states are obviously eigenstates of the self-adjoint
operator (4.6).
Actually, it is not difficult to verify that 
$$
\hat H(\alpha,\varrho)\left|n,p_\perp;\hat\alpha,\varrho\right>=
\left(2n-2\alpha-\varrho p_\perp-{1\over 4}\varrho^2\right)
\left|n,p_\perp;\hat\alpha,\varrho\right>\ .
\eqno(4.13)
$$
It is worthwhile to remark that the key point to obtain the above result is
the fact that the set of states (4.8) is closed with respect to the free action
of the translated degeneracy operators - see eq.s~(4.10). This crucial feature
is peculiar of the set (4.8) and, in particular, does not keep true for the
other complete orthonormal set $\left|n,k;\check\alpha,\varrho\right>$,
which
characterizes the self-adjoint extension of the Hamiltonian with only integer
eigenvalues (like in the "classical" case). This is why the quantum
Hamiltonian (4.11) is the only (essentially) self-adjoint operator with a
continuous non-degenerate spectrum and which commutes with the "Landau-like"
self-adjoint operator (4.6), {\it i.e.} the unique solution of our problem.   

Now, it can be readily checked that the conductance does not change with
respect to the "classical" impurity-free case [7]. As a matter of
fact, starting again from the definition of the 
current operator
$$
\hat J_{1,2}={|e|\hbar\over m\lambda_B}
\hat P_{1,2}(\alpha)\ ,
\eqno(4.14)
$$
in which
$$
\eqalignno{
&\hat P_1(\alpha)=-{1\over \sqrt2}\left[\delta_\varrho(\alpha)
+\bar\delta_\varrho(\alpha)\right]\ , 
&(4.15a)\cr
&\hat P_2(\alpha)={i\over \sqrt2}\left[\delta_\varrho(\alpha)
-\bar\delta_\varrho(\alpha) +{i\over \sqrt2}\varrho\right]\ ,
&(4.15b)\cr}
$$
it immediately follows that, for any normalizable wave packet
$$
\left|n,[f];\hat\alpha,\varrho\right>=\int_{-\infty}^{+\infty}dp_\perp\
f(p_\perp)\left|n,p_\perp;\hat\alpha,\varrho\right>\ ,\quad 
\int_{-\infty}^{+\infty}dp_\perp\left|f(p_\perp)\right|^2=1\ ,
\eqno(4.16)
$$
which describes one electron in the $n$-th conducting band, 
we obtain once again that the average current carried by such a state is
$$
\eqalignno{
&\left<n,[f];\hat\alpha,\varrho|\hat J_{1}|n,[f];\hat\alpha,\varrho\right>
=0\ , &(4.17a)\cr
&\left<n,[f];\hat\alpha,\varrho|\hat J_{2}|n,[f];\hat\alpha,\varrho\right>
=-|e|c{E_H\over B}\ , &(4.17b)\cr}
$$
which shows that the Hall's conductance is always the "classical" one
as in eq.~(2.14), even in the presence of the AB-vortex.
A further important remark is now in order. Taking the limit, when $\alpha$ is
going to zero, of the self-adjoint Hamiltonian (4.11) we recover the standard
impurity-free self-adjoint Hamiltonian (2.6), whose domain is that of regular
wave functions. This means that, if we eliminate the Aharonov-Bohm vortex
potential, contact-interaction is no longer allowed in the presence of a
crossed uniform electric and magnetic fields. Consequently, the analysis of the self-adjoint
extensions of the quantum Hamiltonian does contradict the claim in [3] since no localized bound states are allowed.  
It should also be remarked that the switching on of a weak Hall electric field
does represent a small and smooth perturbation on the system, iff the 
starting unperturbed Hamiltonian is 
$$
\hat\Delta(\alpha)\equiv\sum_{n=0}^\infty
2(n+|\alpha|)\left\{\sum_{k=0}^n\hat P_{n\ge k}(\alpha)
+\sum_{k=n+1}^\infty\hat P_{n<k}(\alpha)\right\}\ .
\eqno(4.18)
$$
Otherwise, for any different choice of the unperturbed Hamiltonian, the
additional electric field cannot represent a smooth perturbation, 
since it involves a change in the domain of the Hamiltonian.

\bigskip
\noindent
{\bf 6.\ Conclusion.}
\medskip
In this paper we have explicitely solved the quantum mechanical {\it
2+1} dimensional problem of the non-relativistic electron in the presence of a
uniform electromagnetic field and of an Aharonov-Bohm vortex potential. The
solution is unique, since it turns out that the quantum Hamiltonian is
essentially self-adjoint in the presence of the uniform electric field. This
is no longer true in the absence of the electric field: in the latter case -
under the assumption of the $O(2)$-symmetry - each radial Hamiltonian allows
for a one-parameter family of self-adjoint extensions. From the explicit
knowledge of the eigenvalues and eigenfunctions of the full Hamiltonian, it is
possible to compute the current and conductance, the results being
the very same as in the "classical" case, {\it i.e.}, in the absence of the
AB-vortex. It should be emphasized that the possibility to obtain the exact
non-perturbative solution is heavily supported by the systematic application of
the algebraic method, which allows to overcome the conflict between the
rotational symmetry in the absence of the electric field and the explicit
symmetry breaking due to the switching on of the uniform electric field itself.

The final result is that the Hall's conductance does not change in the presence
of one impurity described by the AB-vortex. 
The microscopic picture which emerges from the exact solution of the present
one-impurity model can be summarized as follows. In the absence of the
electric field, the general pattern can be described, as it was basically
known [10], by the presence of depopulated integer-valued Landau's levels
and of further real-valued energy levels of finite degeneracy - the
details of this description depending upon the specific self-adjoint
extension of the quantum Hamiltonian, as it was carefully explained in section
3.

On the ground of this model, one is led to the picture of the "broadening" of
the Landau's subbands, owing to the presence of impurities, and to the idea
that the switching on of a weak electric field does basically keep this
feature:
the Hall's conducting states of the electrons should be only those ones lying
within the integer-valued Landau's subbands, the remaining allowed bound states
giving no contribution to the Hall's conductivity.    
On the contrary, the exact solution of the present simple model shows that the
switching on of the uniform electric field drastically changes the above
picture: the electrically splitted Landau's levels are shifted with respect to
the impurity-free case - see eq.s~(2.9) and (4.13) - all the energy
eigenstates belong to the electrically splitted Landau's subbands and the
quantum eigenstates within each subband are necessarily described
by a singular wave function, at variance with the
impurity-free case. It is quite remarkable that, in spite of the above drastic
reshuffling of the quantum states after the switching on of the electric
field, the current and conductance are exactly the same with and
without the AB-vortex. This fact appears to corroborate some deep topological
nature of the Hall's conductance, as it was widely recognized in the
literature [2]. As a matter of fact, the unavoidable presence of singularities
of the eigenfunctions at the vortex position clearly represents the  token of
topological non-triviality.  In other words, since the exact solution of the
one-impurity problem yet necessarily involves singular wave functions at
the impurity position, it means that the underlying configuration manifold in
the general dynamical problem is the punctured plane, which is topologically
non-trivial. This is {\it a fortiori} true in the realistic many impurities
problem, whose exact solution is still unknown.

\bigskip
\noindent
{\bf Acknowledgments}
\medskip
We are very much indebited to Eric Akkermans and St\`ephane Ouvry for
introducing us to the present subject and for quite illuminating discussions,
to Rolf Tarrach for reading the manuscript and for his comments and remarks.
We warmly thank for the hospitality the Department of Physics of the La Plata
University, where the final part of this work has been done and, in particular,
we are grateful to Horacio Falomir and Mariel Santangelo for their interest and
contribution to the comprehension of the mathematical aspects of the present
work. We also thank A. A. Andrianov for useful conversations. This work has
been partially supported by a grant MURST-ex quota 40\%.  

\vfill\eject
\noindent
{\bf References}
\medskip
\item{[1]}K. von Klitzing, G. Dorda and M. Pepper, Phys. Rev. Lett. {\bf
45} (1980) 494.
\item{[2]}A recent review on the subject is: T. Chakraborty and P.
Pietil\"ainen, ``{\it The Quantum Hall Effects}'', Springer-Verlag, Berlin
Heidelberg (1995) and references therein. 
\item{[3]}R. E. Prange, Phys. Rev. B{\bf 23} (1981) 4802; R. Joynt and R. E.
Prange, Phys. Rev. B{\bf 25} (1982) 2943. 
\item{[4]}S. Albeverio, F. Gesztesy, R. Hoegh-Krohn and H. Holden,
``{\it Solvable Models in Quantum Mechanics}'', Springer-Verlag, New York
(1988).
\item{[5]}J. Desbois, S. Ouvry and C. Texier, Nucl. Phys. B{\bf 500} [FS]
(1997) 486.
\item{[6]}Y. Aharonov and D. Bohm, Phys. Rev. {\bf 115} (1959) 485.
\item{[7]}N. Brali\'c, R. M. Cavalcanti, C. A. A. de Carvalho and P. Donatis,
{\tt hep-th/9704091}.
\item{[8]}K. Shizuya, {\tt cond-mat/9710337}.
\item{[9]}A. Feldman and A. H. Kahn, Phys. Rev. B{\bf 1} (1970) 4584.
\item{[10]}A. Comtet, Y. Georgelin and S. Ouvry, J. Phys. A{\bf 22} (1989)
3917.
\item{[12]}I. M. Guelfand and G. E. Chilov, ``{\it Les distributions}'',
Dunod, Paris (1963).
\item{[13]}P. Giacconi, F. Maltoni and R. Soldati, Phys. Lett. B{\bf 441}
(1998) 257.

\vfill\eject

\end
\bye